\documentclass{PoS}
\usepackage{graphicx,epsfig,subfigure,amsmath,tikz,times,mathtools}
\usepackage[subfigure]{ccaption}
\usepackage{slashed}
\title{Strongly Interacting Matter Phase Diagram in the presence of Magnetic Fields in an Extended Effective Lagrangian Approach with Explicit Chiral Symmetry Breaking Interactions}

\ShortTitle{
Strongly Interacting Matter PD in the presence of \emph{B} in an Ext.-ELA with Explicit $\chi$ Simmetry Breaking Interactions
}

\author{\speaker{J. Moreira},$^a$ J. Morais,$^a$ B. Hiller,$^a$ A. A. Osipov $^b$ and A. H. Blin $^a$ 
				\\
				\llap{$ˆa$} CFisUC, Department of Physics, University of Coimbra,\\
				P-3004-516 Coimbra, Portugal\\
				\llap{$ˆb$} JINR - Bogoliubov Laboratory of Theoretical Physics\\
				141980 Dubna, Moscow region, Russia\\
        E-mail: \email{jmoreira@fis.uc.pt}}

\abstract{
Extensions of the NJL model which go beyond the original 4-quark interaction, which drives the dynamical mass generation, have proven to be quite successful in describing low energy hadronic phenomenology. The inclusion of 8-quark interaction terms solved a metastability problem of the effective potential
introduced by the inclusion of the 6-quark 't Hooft determinant term in the 3-flavor version of the model (needed to eliminate the unwanted U(1) axial symmetry) . This model, that has proven to be quite powerful and feature-rich, has been expanded to include all the spin-0 terms, without and with explicit chiral symmetry breaking, which are of the same order as the 't Hooft flavor determinant in a 1/Nc expansion resulting in an unprecedented success in reproducing
the low lying scalar and pseudoscalar meson spectra. This success can be seen as a result of the inclusion of the full chiral symmetry breaking pattern. The two critical endpoints which are obtained in the temperature/chemical potential phase diagram are shifted to lower chemical potential and higher temperature when the effect of magnetic field is taken into account. For the studied magnetic field strengths (in the range $eH=0-0.4~GeV^2$) no significant extra transitions are seen to appear.
}

\FullConference{XVII International Conference on Hadron Spectroscopy and Structure\\
                 25-29 September, 2017\\
                 University of Salamanca, Salamanca, Spain}

\begin{document}
\section{Introduction}

The influence of magnetic fields in the Phase Diagram has been under intense scrutiny due to their relevance for instance in the context of Heavy Ion Collisions, compact stars and early Universe phases \cite{Andersen:2014xxa}. Due to the limitations encountered by lattice QCD studies at finite chemical potential the insight given by effective models can be extremely helpful. Here we will focus on results obtained with the model introduced in \cite{Osipov:2012kk, Osipov:2013fka}.

\section{Model}
The thermodynamical potential can be obtained from the model lagrangian using standard techniques and is given by (for more details see \cite{NJLH8qPD_PRD81_2010,NJLH8qmPD_PRD91_2015} and references therein): 
\begin{align}
\Omega=&\mathcal{V}_{st}+\sum_i\frac{N_c}{8\pi^2}\left(J_{-1}\left[M_i,T,\mu_i\right]+C\left[T,\mu_i\right]\right)\\	
\mathcal{V}_{st}=&\frac{1}{16}\bigg(4 G \left(h_i^2\right) + 3 g_1 \left(h_i^2\right)^2 + 3 g_2 \left(h_i^4\right) + 4 g_3 \left(h_i^3 m_i\right)
			+ 4 g_4 \left(h_i^2\right) \left(h_j m_j\right) + 2 g_5 \left(h_i^2 m_i^2\right)  \nonumber\\ 
	    &+ 2 g_6 \left(h_i^2 m_i^2\right)+ 4 g_7 \left(h_i m_i\right)^2 + 8 \kappa h_u h_d h_s -8 \kappa_2 \left( m_u h_d h_s + h_u m_d h_s +  h_u h_d m_s\right)\bigg)\bigg|^{M_i}_0\nonumber\\
\Delta_f=&  M_f-m_f\nonumber\\ 
        =& -G h_f -\frac{g_1}{2} h_f(h_i^2) - \frac{g_2}{2} (h_f^3) - \frac{3 g_3}{4} h_f^2 m_f\nonumber\\
				 &-\frac{  g_4}{4} \left( m_f \left(h_i^2\right)+2 h_f(m_i h_i)\right) - \frac{g_5 + g_6}{2} h_f m_f^2
				- g_7 m_f (h_i m_i) - \frac{\kappa}{4} t_{fij}h_i h_j - \kappa_2 t_{fij}h_i m_j\nonumber\\				
J_i\left[M\right]=&16\pi^2\Gamma (i+1)\!\int  \frac{\mathrm{d}^4p_E}{(2\pi)^4}\,\hat\rho_\Lambda \frac{1}{(p_E^2+M^2)^{i+1}}%\nonumber\\ 
								 =
								%&
								16\pi^2\!\int  \frac{\mathrm{d}^4p_E}{(2\pi)^4} \int^\infty_0 \mathrm{d}\tau \,\hat\rho_\Lambda\tau^{i}\mathrm{e}^{-\tau (p_E^2+M^2)}. \nonumber\\
J_{i+1}\left[M\right]=&-\frac{1}{2M}\frac{\partial}{\partial M}J_{i}\left[M\right]\nonumber
\end{align}
Here $h_i$ ($i=u, d, s$) are proportional to the quark chiral condensates. Their value is determined by the minimization of the thermodynamical potential. The quark-loop integrals $J_i$ are rendered finite by using a regularization kernel which corresponds to two Pauli-Villars subtractions in the integrand $\hat\rho_\Lambda\left(\tau\Lambda^2\right)=1-(1+\tau\Lambda^2)\mathrm{e}^{-\tau\Lambda^2}$. The dynamical and current mass of the quarks is denoted respectively by $M$ and $m$.

The effect of a static homogeneous magnetic field, $H$, can be modeled through the introduction of Landau level quantization. In the present case this amounts to the substitution of the integration in transverse momenta (with respect to the direction of the magnetic field which we take to be $z$) by a sum over discrete values:
$  \int\frac{\mathrm{d}^2p_\perp}{\left(2\pi\right)^2}
  \rightarrow \frac{2\pi\left|q\right|H}{\left(2\pi\right)^2} \frac{1}{2}\sum_{s=-1,+1}\sum_{l=0}^{+\infty}$, with the momentum substituted by 
 	$p^2_\perp\rightarrow (2 l+1-s)\left|q\right| H$, where $q$ is the electric charge for each quark.

\section{Results}

\begin{figure}
\centering
	\subfigure{\includegraphics[width=0.32\textwidth]{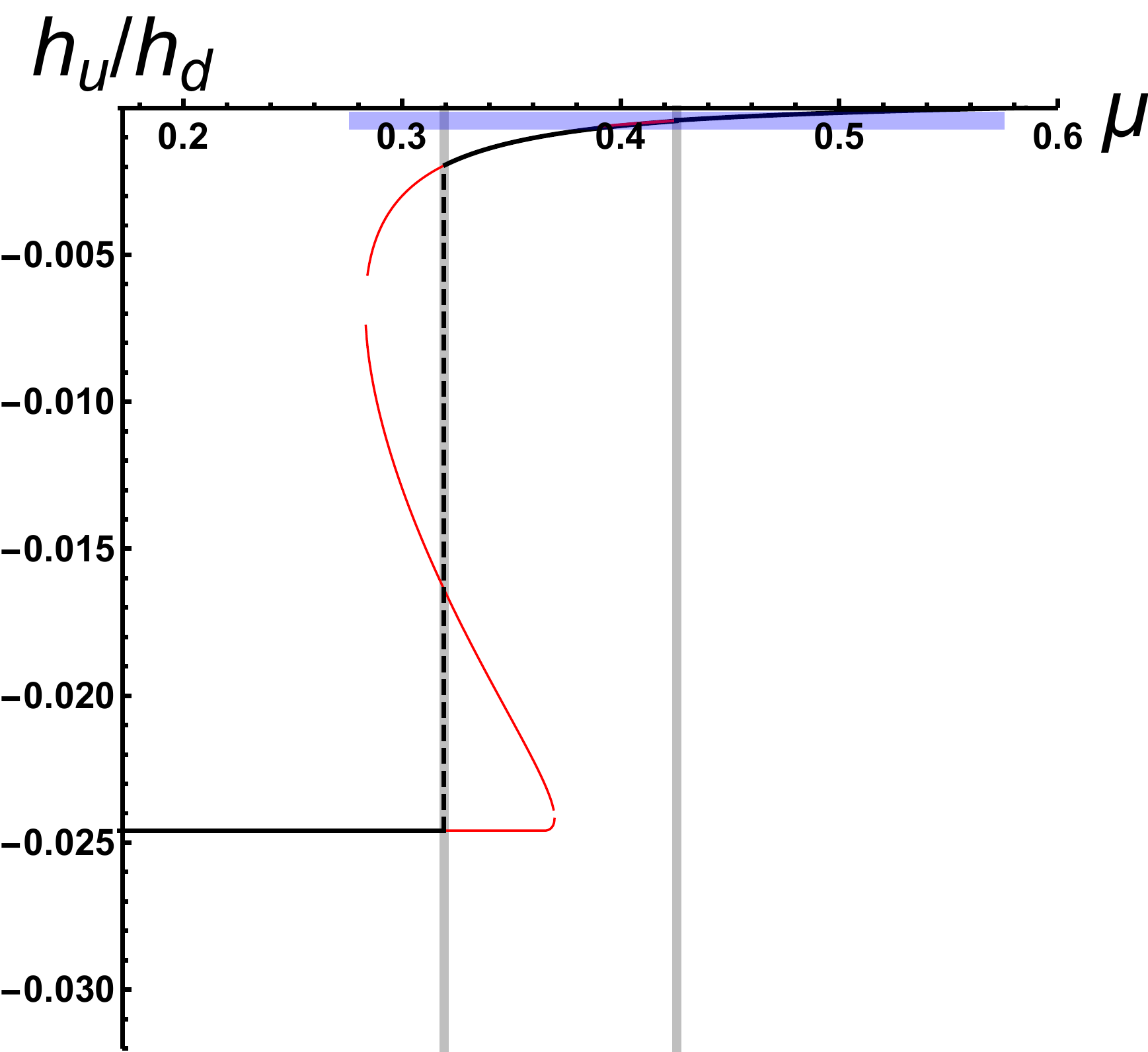}}
	\subfigure{\includegraphics[width=0.32\textwidth]{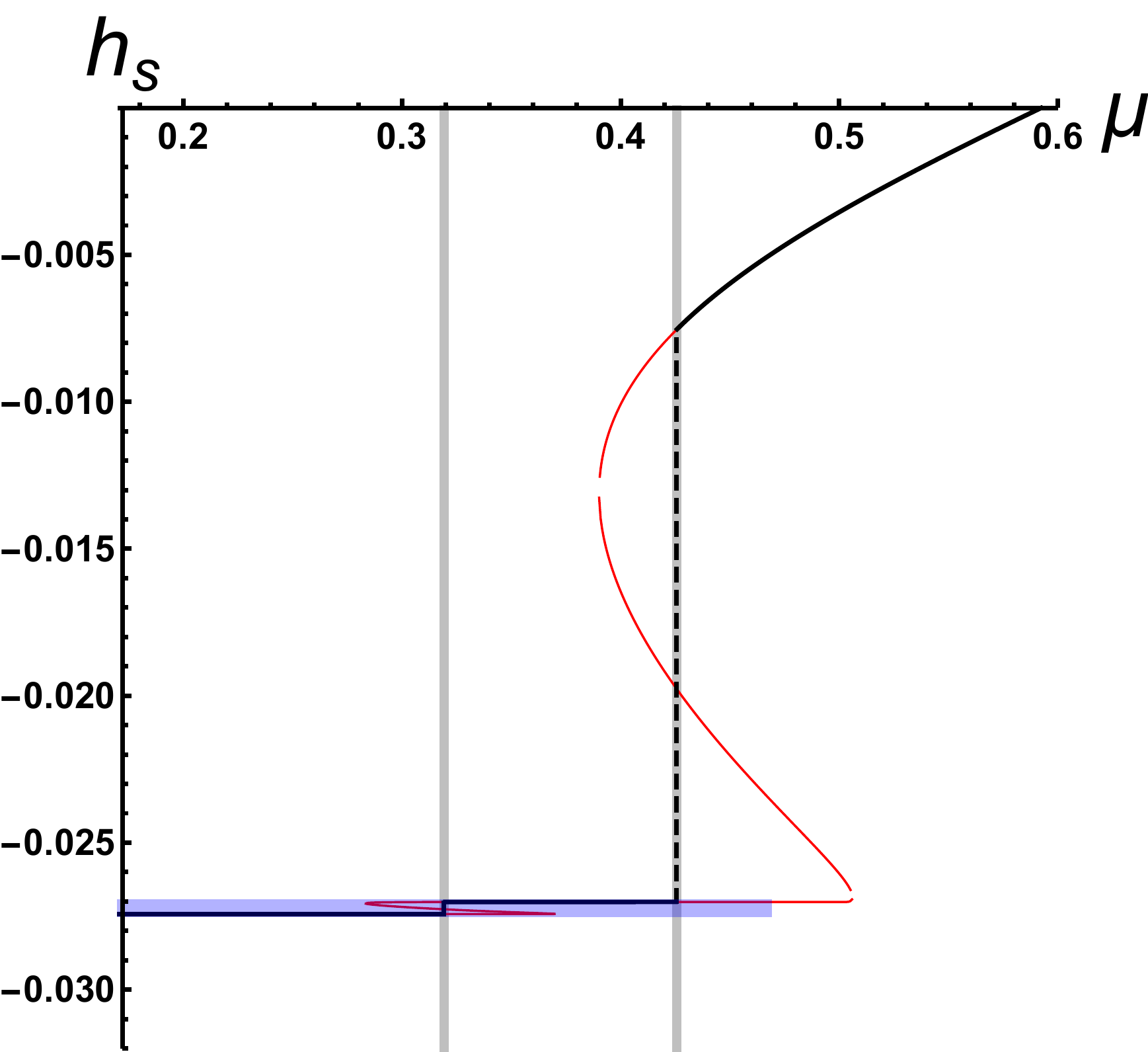}}
\\
	\subfigure{\includegraphics[width=0.32\textwidth]{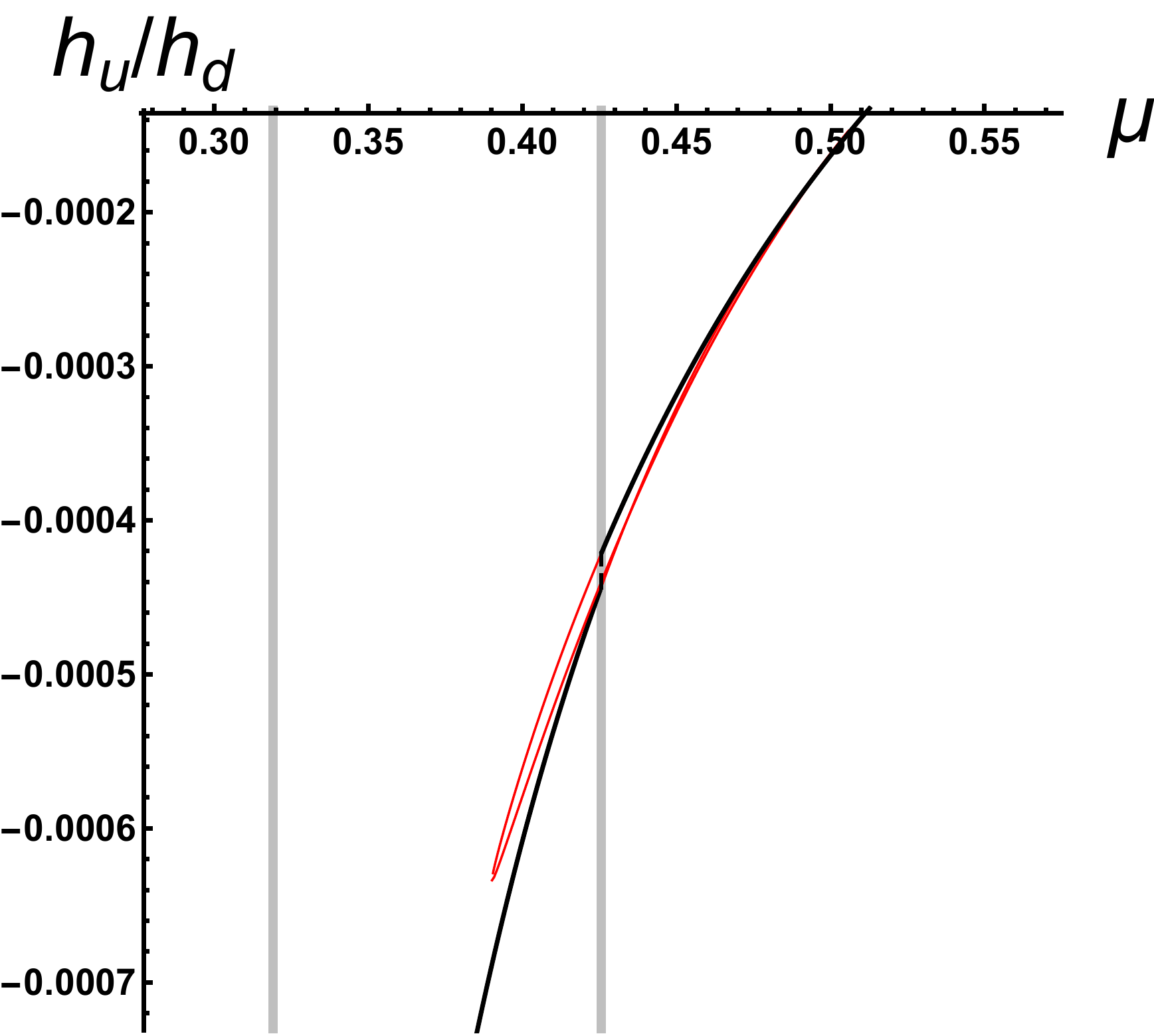}}
	\subfigure{\includegraphics[width=0.32\textwidth]{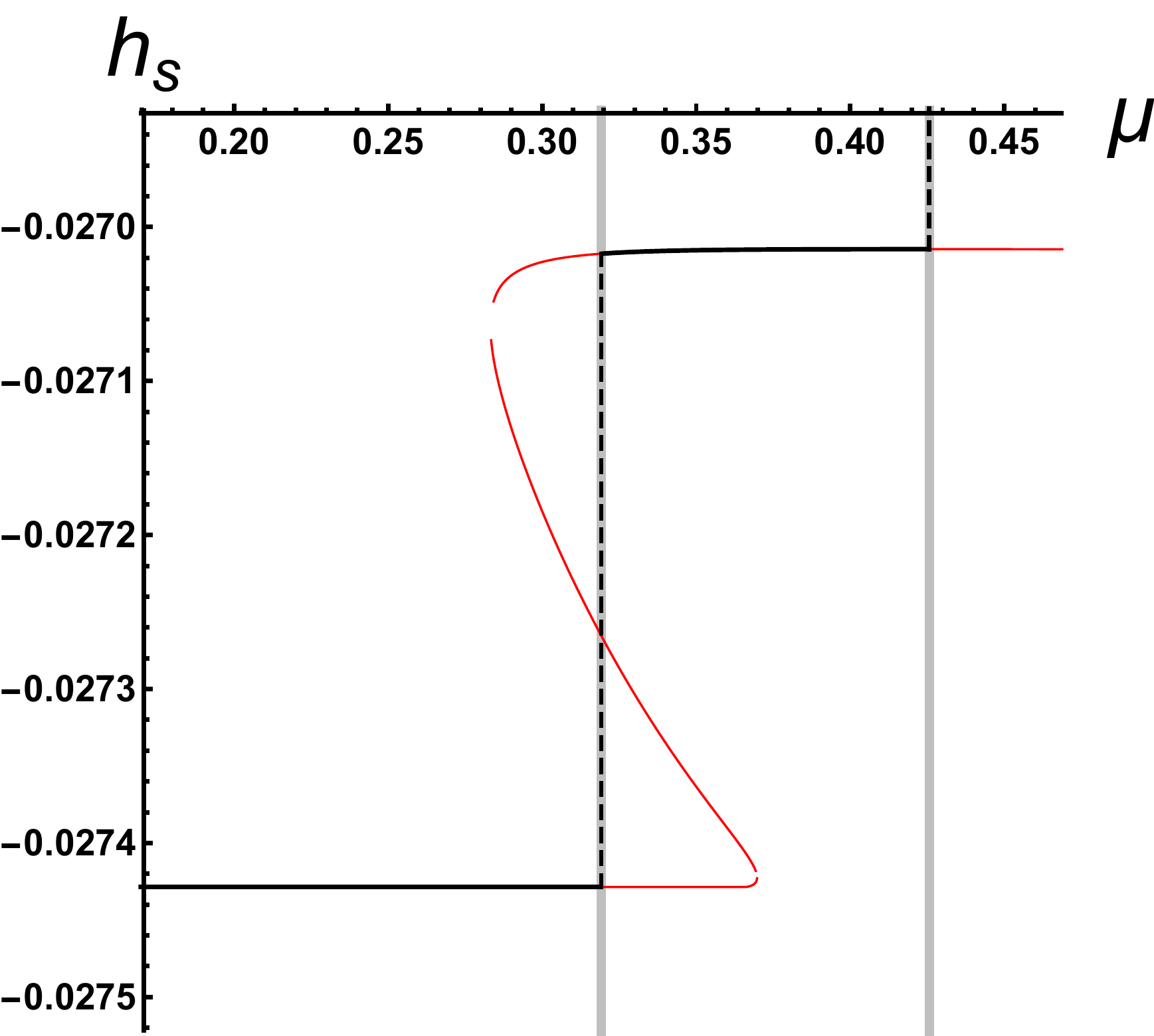}}
\caption{
\label{grafhiT0Hq0000} 
Quark condensates satisfying the gap equations as functions of chemical potential for vanishing temperature and no magnetic field. The two first order transitions are marked by the vertical grey bands. Black denotes the solution which minimizes the thermodynamical potential. The bottom row corresponds to zooms on the highlighted area in the upper row. A much larger jump for the light quark condensate occurs in the transition occurring at lower chemical potential whereas a much larger jump in the strange quark chiral condensate occurs in the second transition.}
\end{figure}

Details about the choice of model parameters can be found in \cite{NJLH8qmPD_PRD91_2015}: the quark current masses are fixed to $m_u=m_d=4~\mathrm{MeV}$, $m_s=100~\mathrm{MeV}$, the coupling constants, $G$, $g_i~(i=1,\ldots,8)$, $\kappa$ , $\kappa_2$, and the cutoff, $\Lambda$, are fitted as to reproduce several properties of the low lying meson spectra. The study of the corresponding phase diagram revealed the existence of two first order transitions lines ending each in a critical endpoint. The transition occuring at a lower chemical potential/temperature can be associated with a large jump in the light quarks chiral condensate and the other with a jump in the strange quark condensate. These jumps can be seen in Fig. \ref{grafhiT0Hq0000}.

\begin{figure}
	\subfigure{\includegraphics[width=0.32\textwidth]{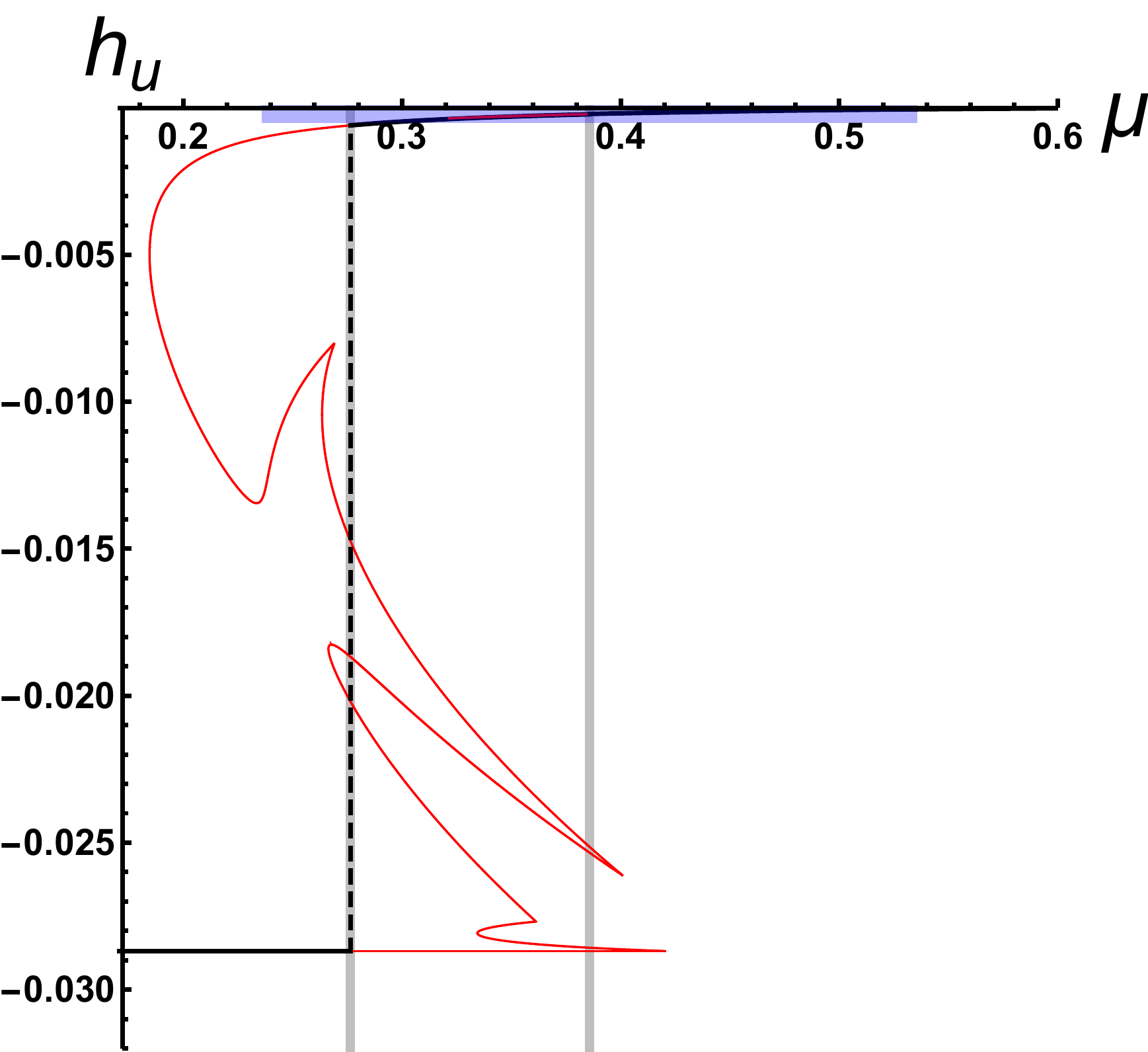}}
	\subfigure{\includegraphics[width=0.32\textwidth]{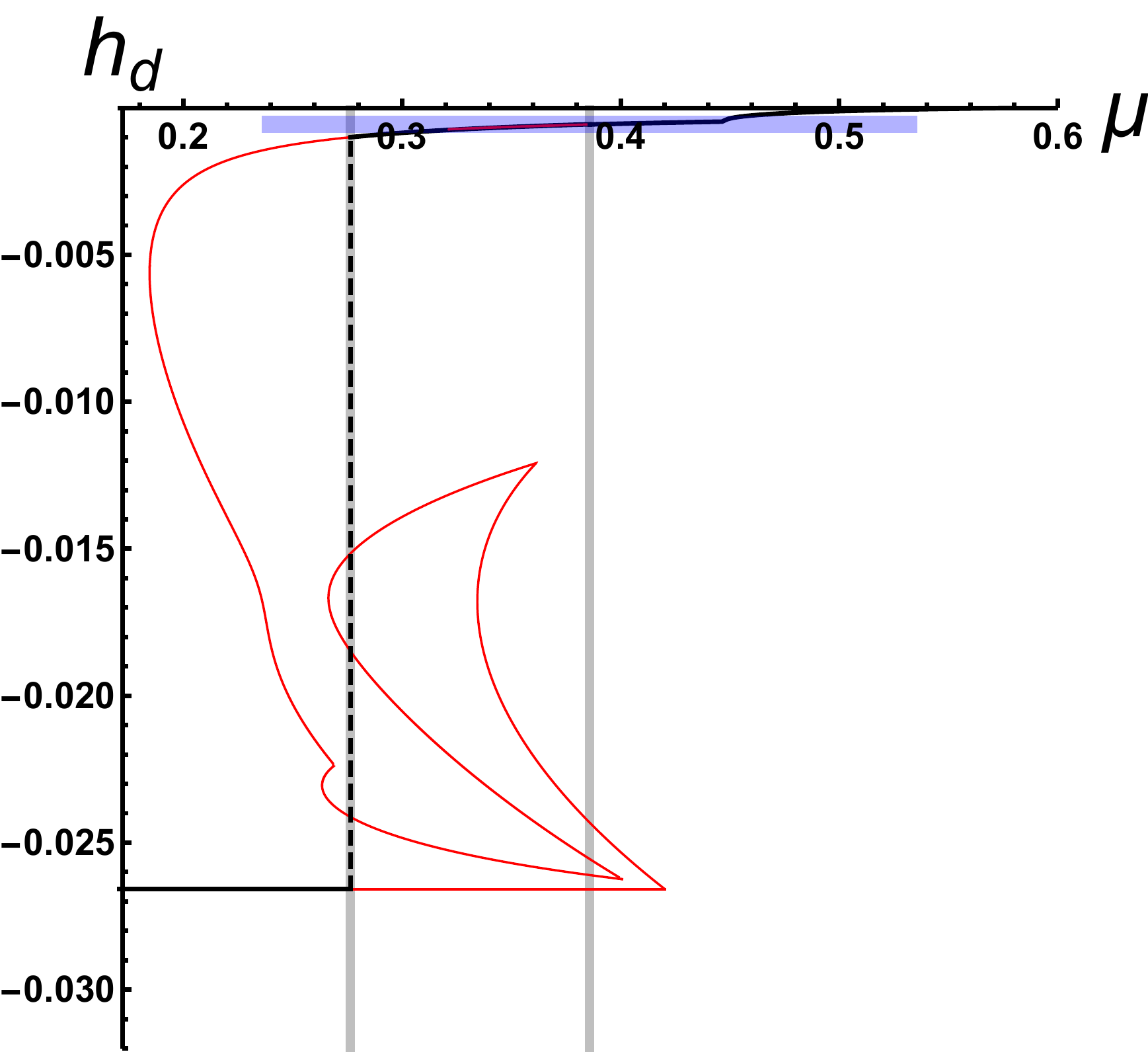}}
	\subfigure{\includegraphics[width=0.32\textwidth]{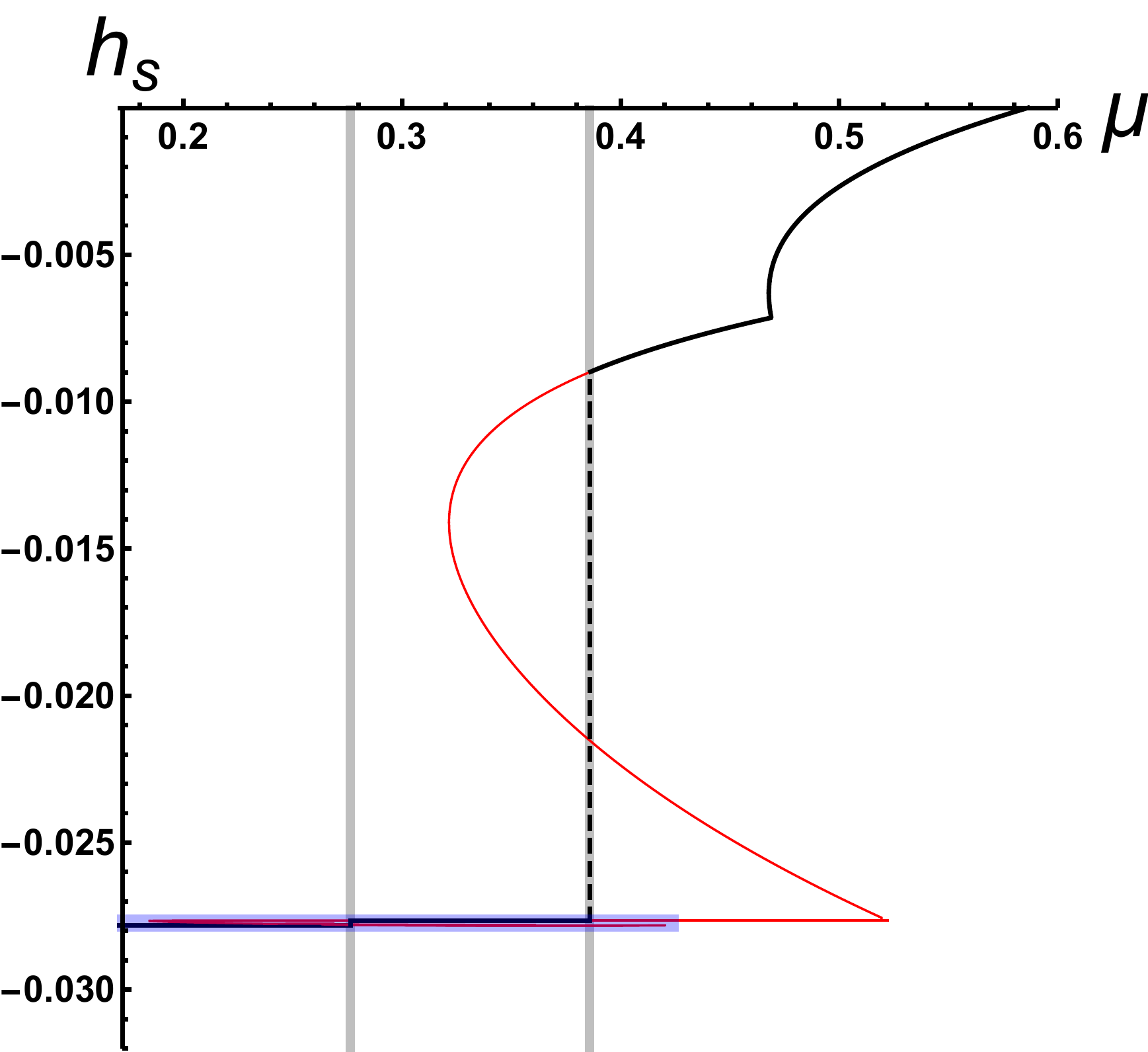}}
\\
	\subfigure{\includegraphics[width=0.32\textwidth]{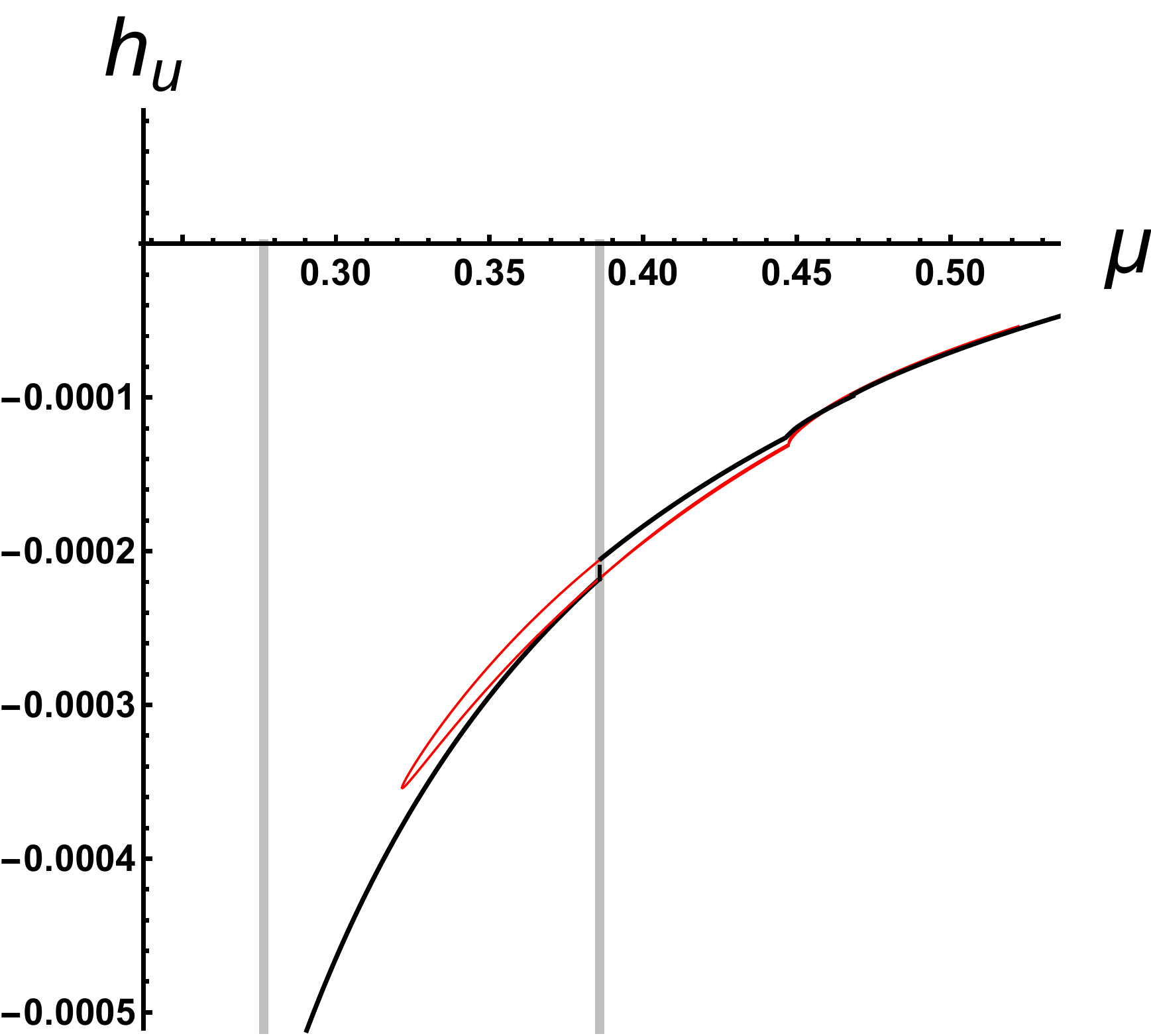}}
	\subfigure{\includegraphics[width=0.32\textwidth]{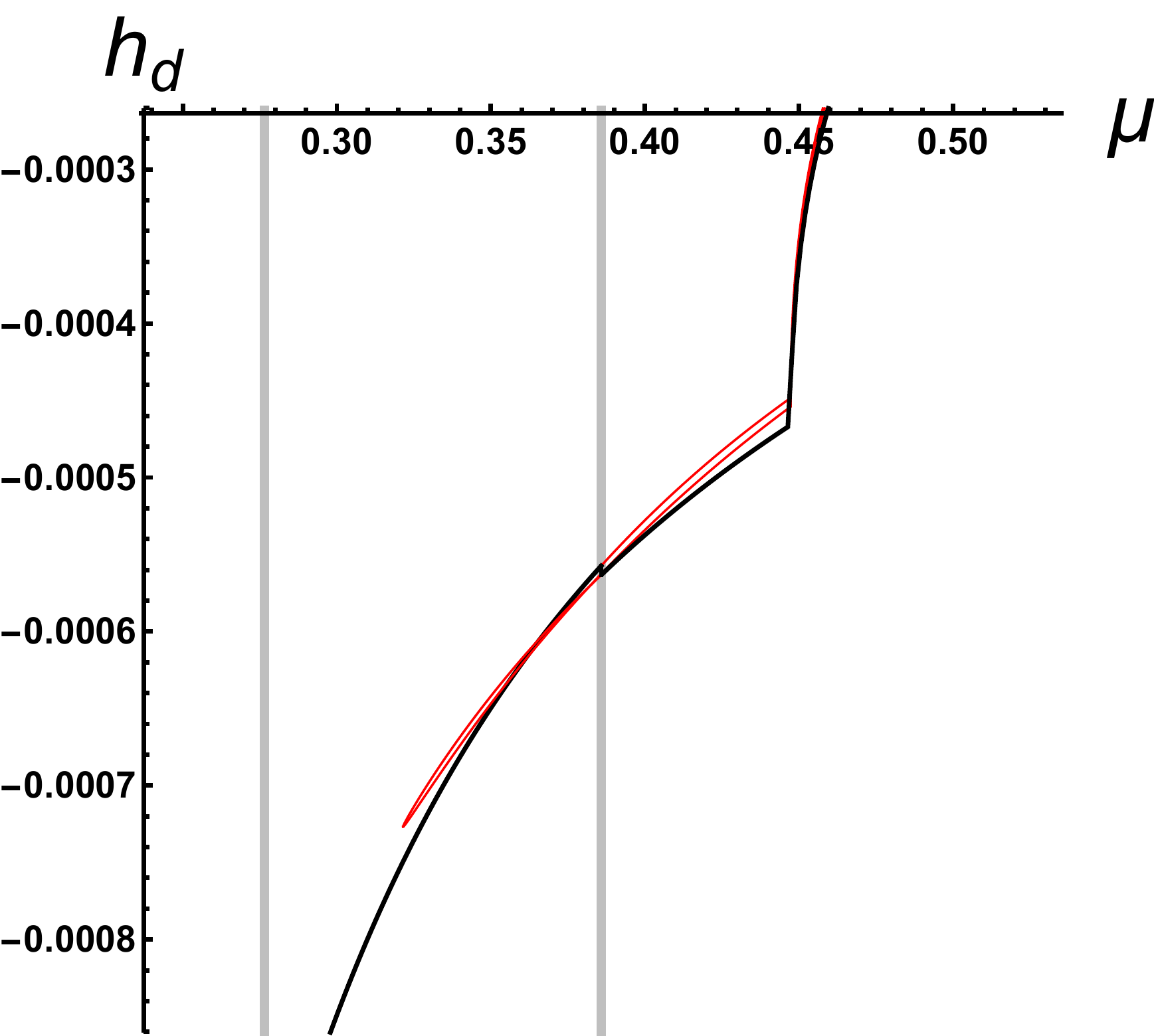}}
	\subfigure{\includegraphics[width=0.32\textwidth]{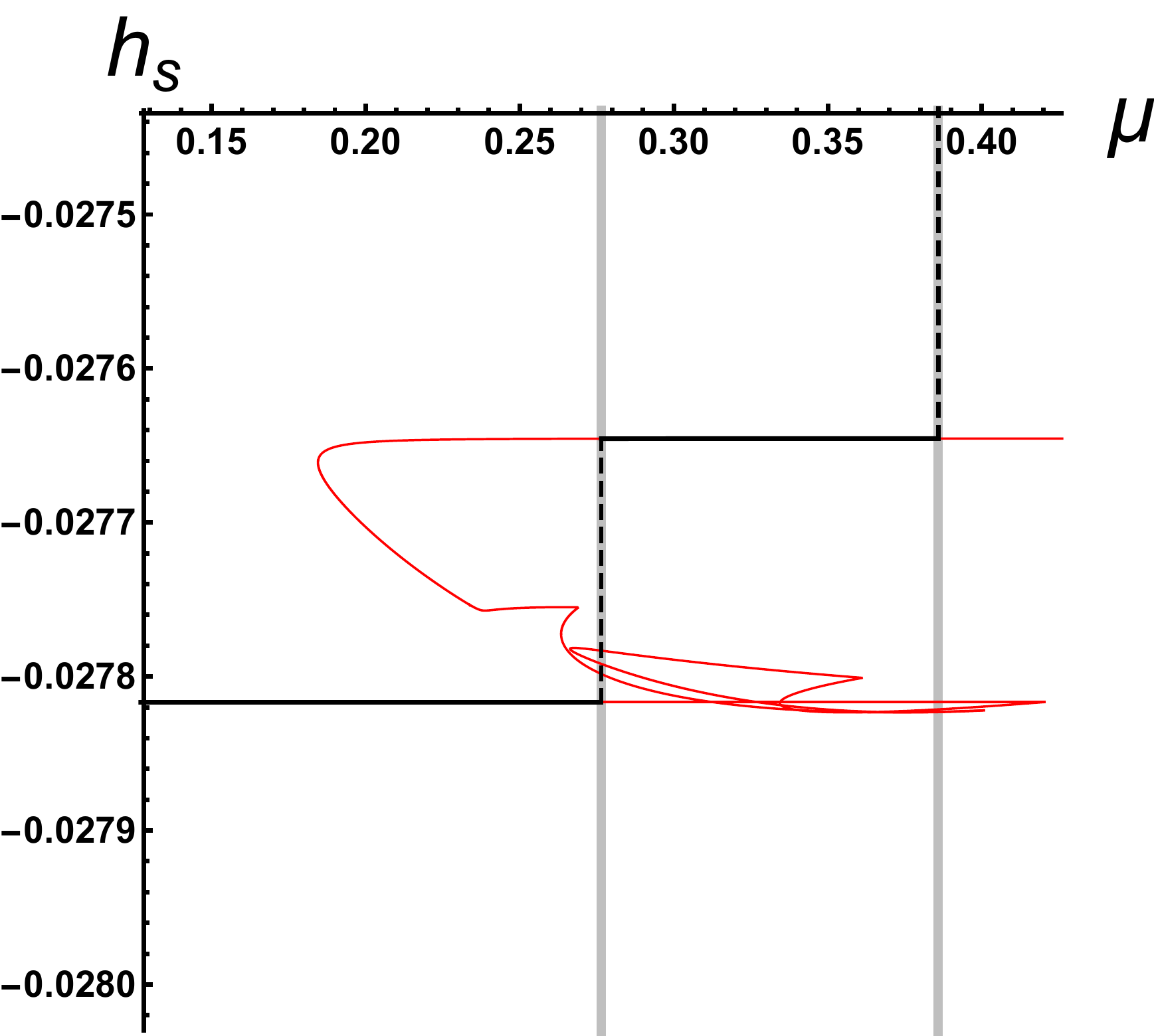}}
\caption{
\label{grafhiT0Hq0300} 
Quark condensates satisfying the gap equations as functions of chemical potential for vanishing temperature and a finite magnetic field $e H=0.3~GeV^2$. The solution branches are distorted by the inclusion of finite $H$ but the jump (the physical solution is represented in black) bypasses the most severe deformations and no new significant transitions are introduced for moderate values of $H$. Bottom row: zooms as in Fig. \ref{grafhiT0Hq0000}.}
\end{figure}

The inclusion of a finite magnetic field deforms the solution branches as can be seen in Fig. \ref{grafhiT0Hq0300} and lifts the degeneracy between $u$ and $d$ (in the present study their current mass is considered equal) due to the difference in electric charge. The kinks in the solution branches are due to the Landau level quantization. 

\begin{figure}
\centering
	\subfigure{\includegraphics[width=0.32\textwidth]{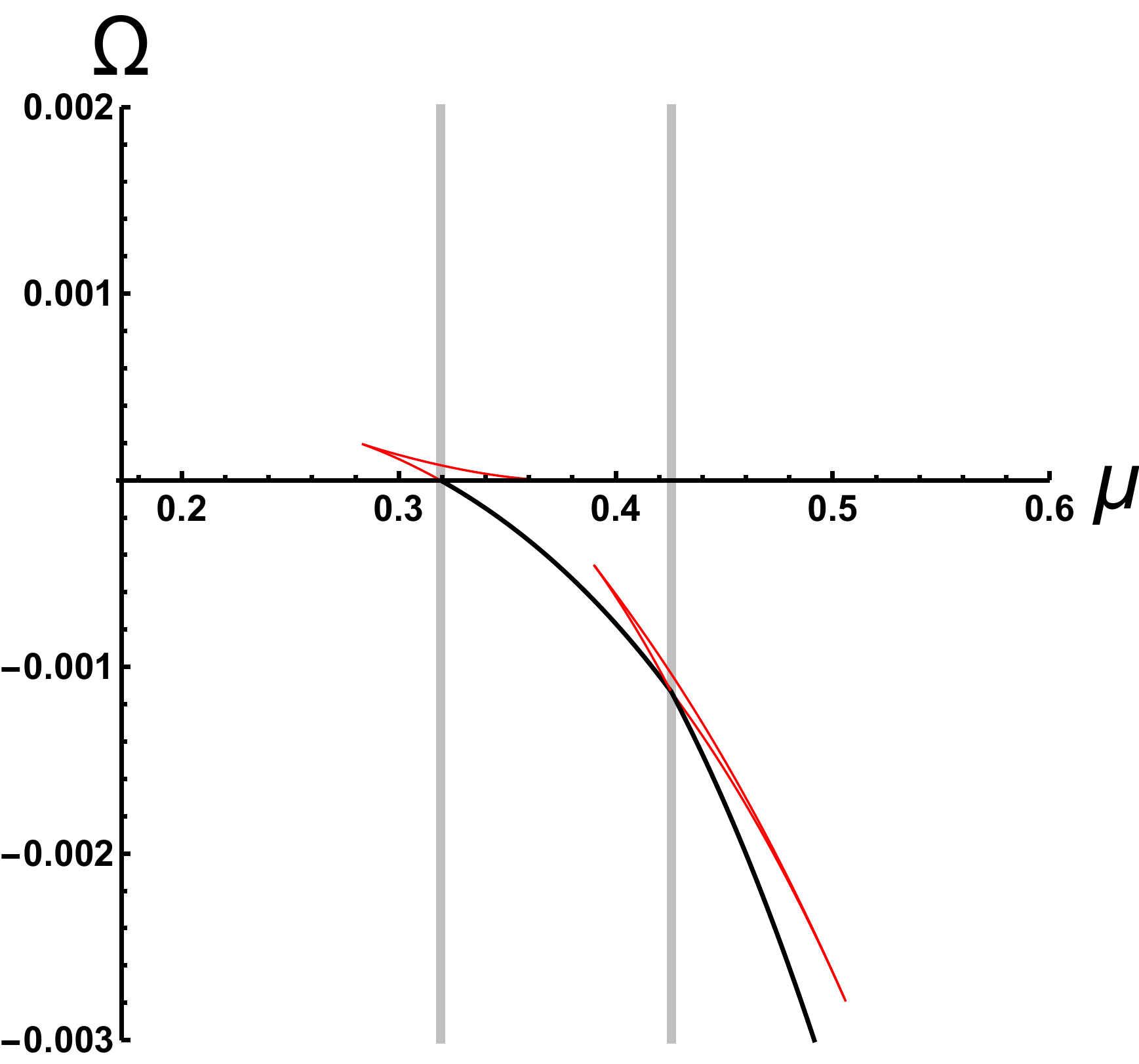}}
	\subfigure{\includegraphics[width=0.3\textwidth]{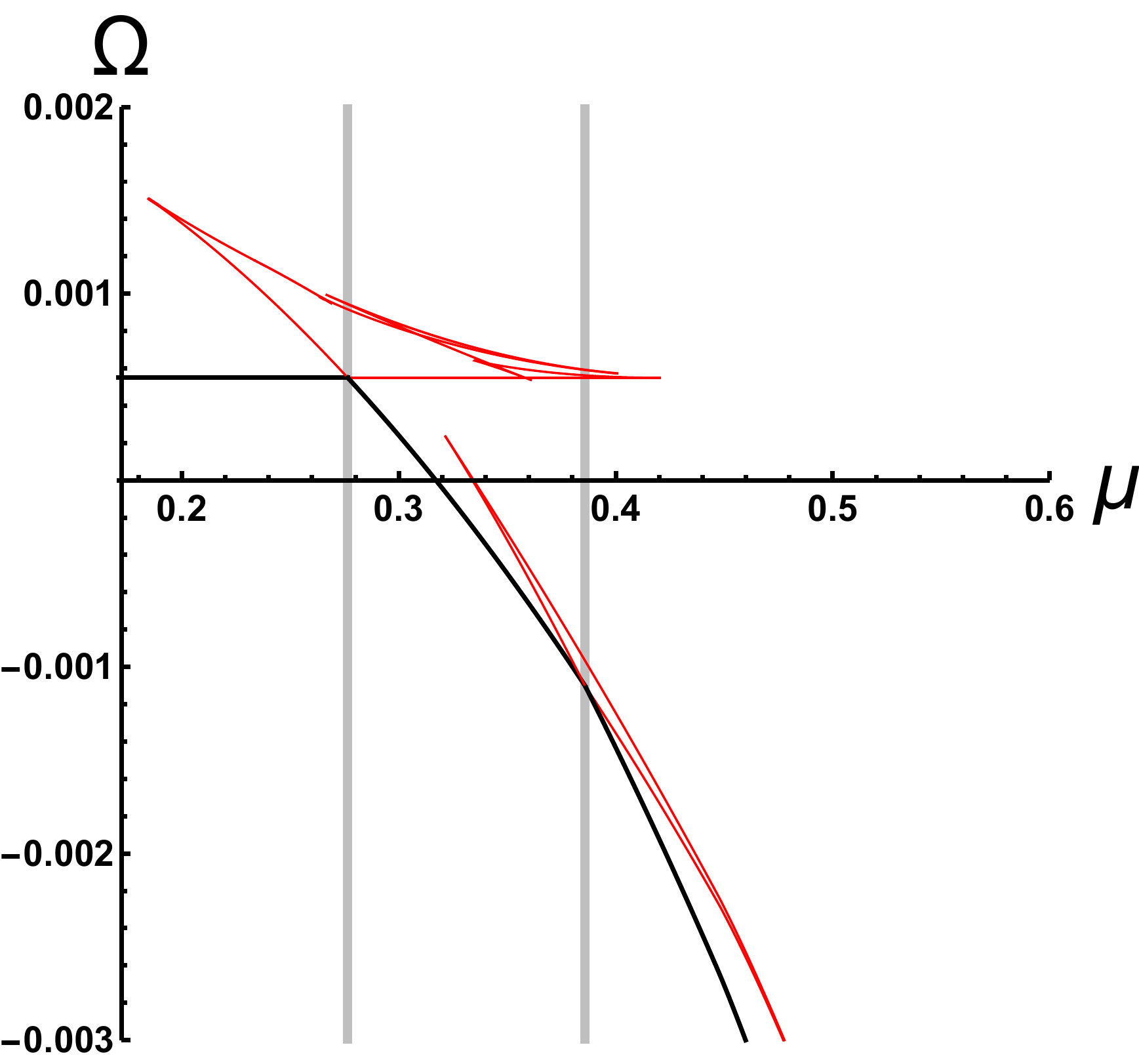}}
\caption{\label{grafPotT0Hqxxx}Thermodynamical potential as a function of chemical potential without and with a magnetic field. On the left-hand side $e H=0$ and on the right-hand side $e H=0.3~GeV^2$.}
\end{figure}

\begin{figure}
\centering
	\subfigure{\label{PDHqxxx}\includegraphics[width=0.43\textwidth]{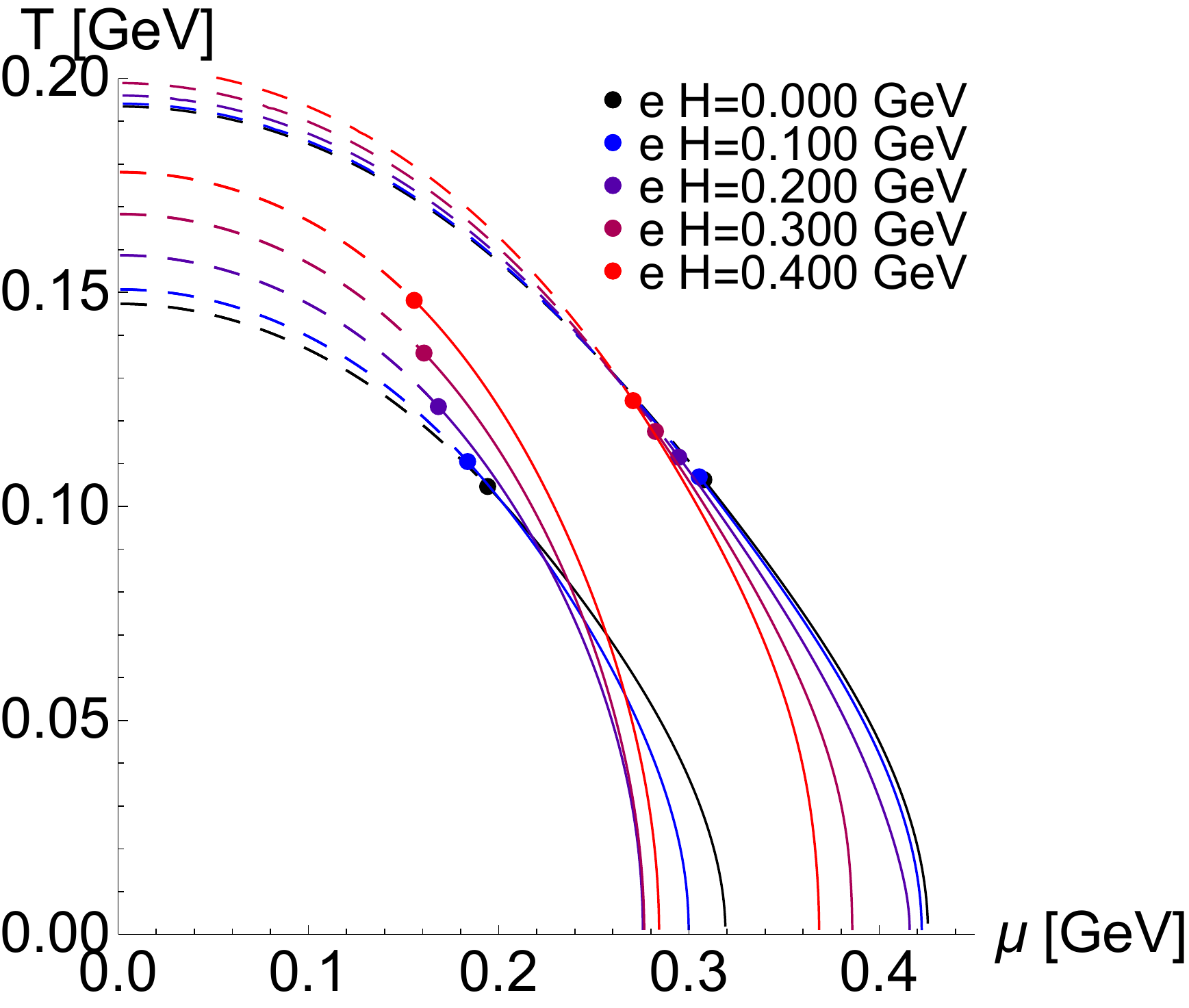}}
	\subfigure{\label{grafmucritvsHVar}\includegraphics[width=0.43\textwidth]{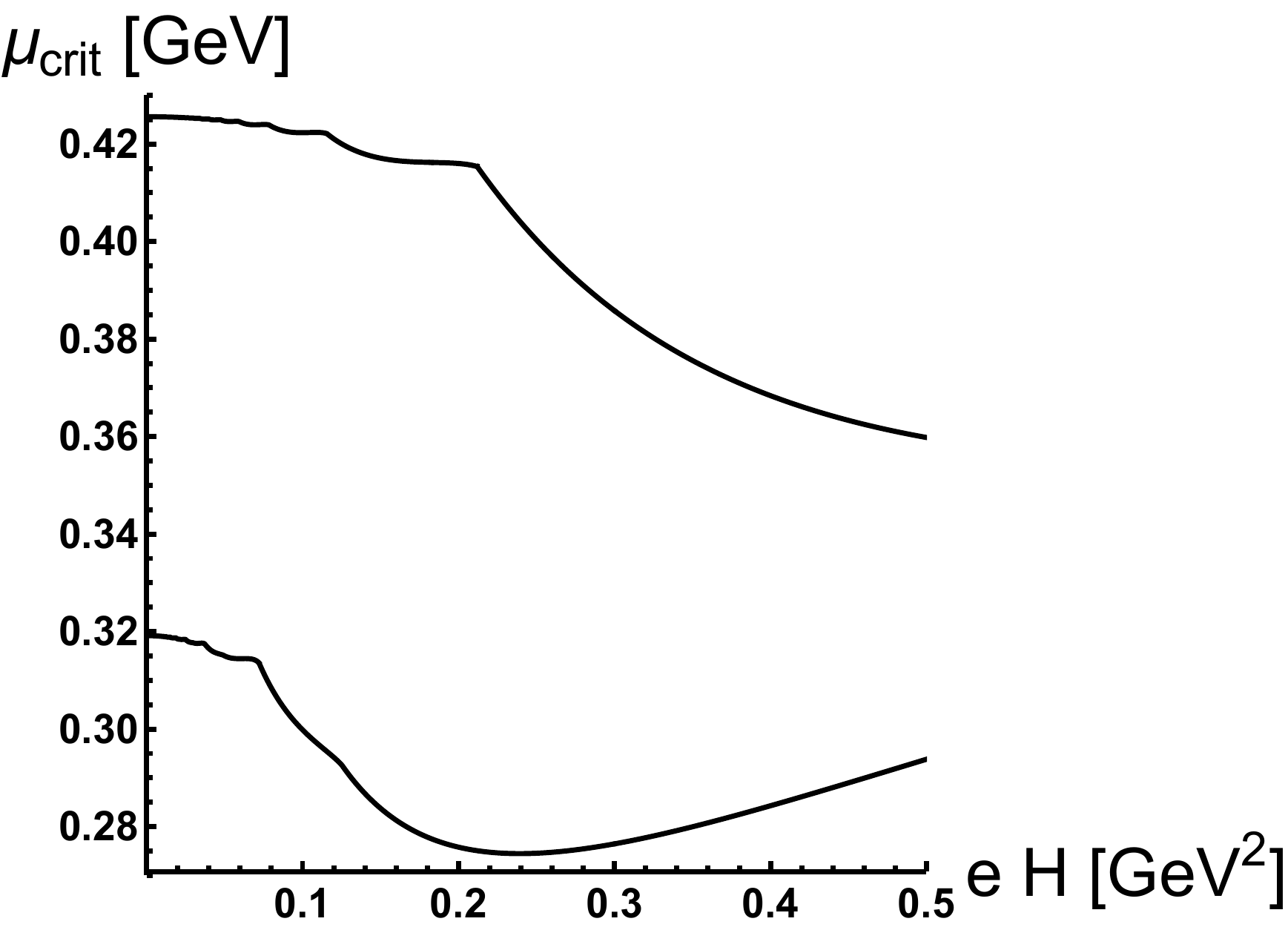}}
	\caption{Left-hand side: phase diagram for several magnetic field strengths. Right-hand side: magnetic field dependence of the critical chemical potential for the first order transitions at vanishing temperature.}
\end{figure}

Despite this deformation of the solution branches most of its effect is \emph{bypassed} by the first order transition jumps and we only see a shift in the transition location with no new transitions being introduced (see Fig. \ref{grafPotT0Hqxxx}). The shift in the critical endpoints with increasing magnetic field strength, which is more pronounced for the transition associated with the lighter quarks, can be seen in Fig. \ref{PDHqxxx}. The magnetic field strength dependence of the chemical potential for the first order transitions at vanishing temperature can be seen in Fig. \ref{grafmucritvsHVar}. In the case of the transition associated with the larger jump in the light quark condensate a non-monotonic behavior is observed as the initial decrease is followed by an increase for larger magnetic field strength.

\section{Conclusions}

We have analyzed the influence of a magnetic field in the phase diagram of strongly interacting matter using a recently developed model lagrangian which includes  all spin-0 interaction terms at leading and sub-leading order in $\frac{1}{N_c}$ counting using a covariant regularization which is kept consistently on both vacuum and medium contributions. A shift in position of the critical endpoints towards lower chemical potential and higher temperature occurs with increasing magnetic field. No substantial new transitions as the ones reported in \cite{PCosta} appear in the phase diagram. The possibility that these are an artifact of the regularization procedure (in other works the medium contribution is usually considered unregularized) should be investigated. 

\section*{Acnowledgments}
This work was supported by CFisUC and the Funda\c{c}\~{a}o para a Ci\^{e}ncia e Tecnologia under project No. UID/FIS/04564/2016 and grants SFRH/BPD/110421/2015, SFRH/BD/110315/2015.

\end{document}